\documentclass[aps,prl,twocolumn,superscriptaddress]{revtex4}

\usepackage{graphicx}
\usepackage{epsf}
\usepackage{amssymb}
\begin{document}
\title{ Plasmonic  enhanced lasers of nano-colloidal  fluids}
\vspace{0.5cm}
\author{ N. Ghofraniha$^\ast$}
\affiliation{IPCF-CNR, UOS Roma Kerberos, University La Sapienza, P.le A. Moro 5, 00185, Roma, Italy}
\author{P. Andr\'{e}$^\ast$}
\affiliation{School of Physics and Astronomy (SUPA), University of St Andrews, St Andrews, UK}
\affiliation{RIKEN Advanced Science Institute, Hirosawa 2-1, Wako, Saitama 351-0198, Japan}
\author{A. Di Falco}
\affiliation{School of Physics and Astronomy (SUPA), University of St Andrews, St Andrews, UK}
\author{C. Conti}
\affiliation{Department of Physics, University La Sapienza, P.le A. Moro 5, 00185, Roma, Italy}
\affiliation{Istitute for Complex-Systems CNR, UOS Sapienza, University La Sapienza, P.le A. Moro 5, 00185, Roma, Italy}

\begin{abstract}
\noindent {\bf   ABSTRACT: 
Localized surface plasmon resonances have recently attracted considerable attention due to their ability to 
dramatically enhance near-field optical intensities and boost nanoscale light-matter interactions. 
Here we demonstrate unambiguously that polyhedral silver nanoparticles can be tailored to promote enhanced coherent emission from organic dye 
dispersions in a cavity. In particular, we observe that nanoparticles supporting resonances close to the absorption peak 
of the dye yield to more efficient lasing, up to one order of magnitude, which is also evidenced by the narrowing of the 
emission spectral line down to 3 nm. Moreover we observe emission saturation and energy dependent spectral shifts associated with Kerr effects, 
which demonstrate the general relevance of non-resonant nonlinearities in plasmonic enhanced laser emissions.
Works in this area will contribute to the design of novel kind of laser sources providing ultra-short pulsed operation and 
structural tunability.}\\
 
\end{abstract}

\maketitle 
\noindent

Understanding the fundamental light-matter interactions at a scale much smaller than the wavelength is pivotal to any effort to reliably engineer laser emission. 
This relies on controlling frequency and bandwidth \cite{Landau}, to ultimately design ultra-short laser pulses. 
Localized surface plasmons (LSP) can be optically excited and are associated with sustained oscillations of free electrons localized 
at the metal/dielectric interface. \cite{Bar03,Zay03} LSP oscillations are coherent and accumulate energy from the external excitation.\\
These LSP can be tuned with the size and the shape of metallic structures which can be fabricated through a wide range of approaches 
including lithography and wet-chemistry.\cite{For12,Vol09,DiF10,Rod09,Sch12,Lin99,Jai06} Such nanometer scale sources of intense coherent electromagnetic fields can enhance 
inherently weak physical processes, such as fluorescence and Raman scattering of single molecules with potential applications ranging from 
biomedicine to lithography, and from microscopy to information technology.\cite{Maier2003,Ank03,Pil11,Ber12,Plo12,Lin12,Oul12,Yin12,Kab09,Liu11NM,DiF11} 

\begin{figure}[h]
\includegraphics[width=0.45\textwidth]{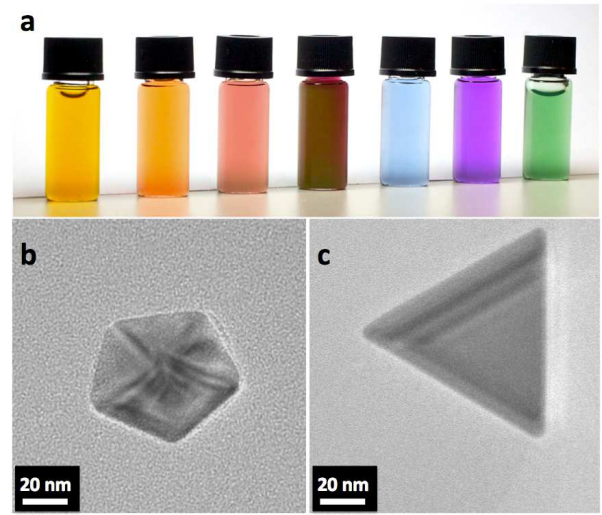}
\caption{ {\bf (a)} Picture of selected samples;  {\bf(b,c)} TEM images of silver nanodecaheadrons and nanoprisms (scale bars are 20 nm). 
}
\label{fig1}
\end{figure}

\begin{figure*}
\includegraphics[width=0.9\textwidth]{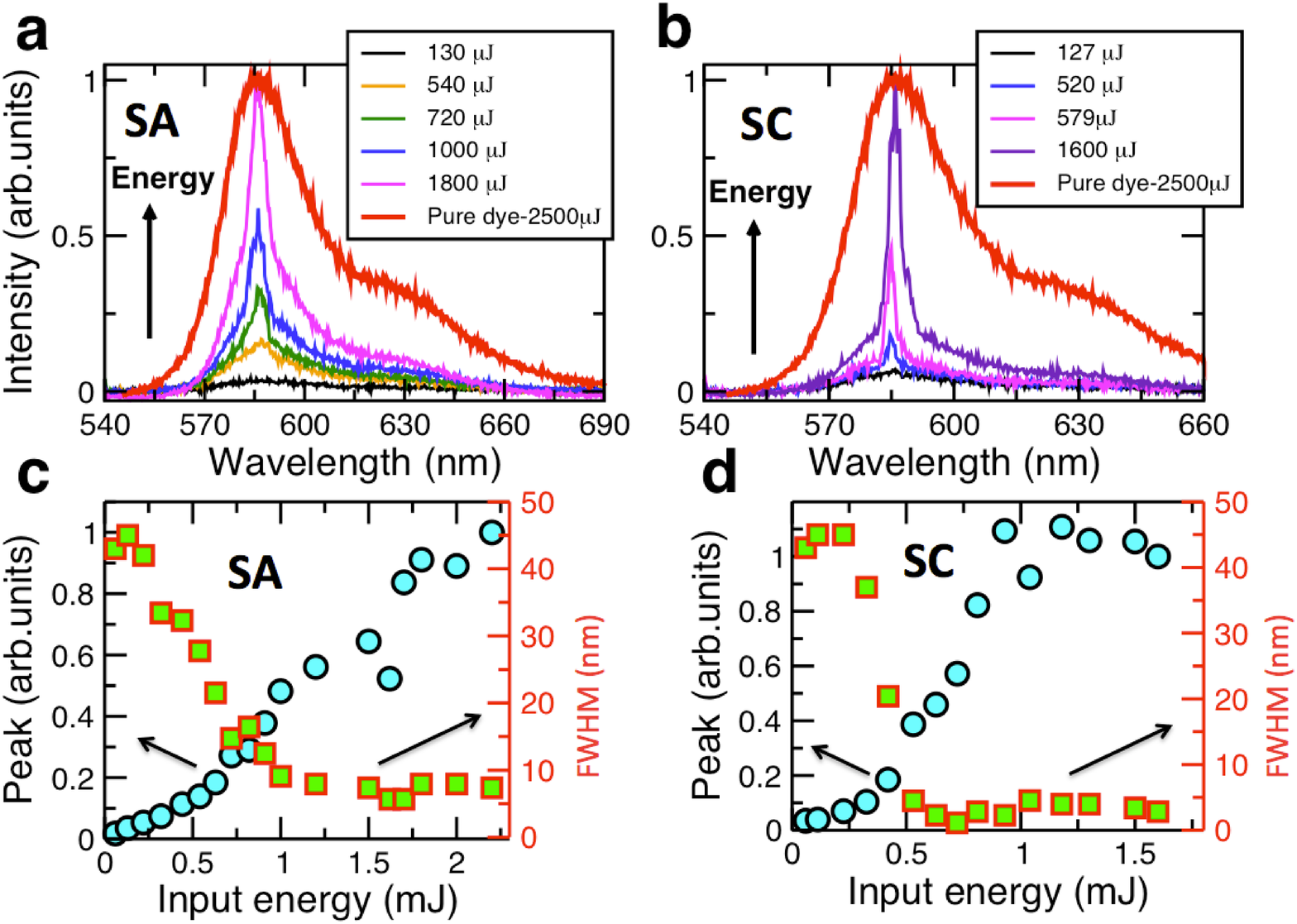}
\caption{ {\bf (a)} Emission spectra at increasing pump energy of samples SA (nanospheres) and 
{\bf(b)} SC (nanodecaheadrons); thick red line: emission of pure dye solution at high pump energy. 
{\bf(c)} spectral width calculated as full width half maximum (FWHM) ($\blacksquare$) and peak intensity ({\Large$\bullet$}) of the emission, for sample SA when increasing input energy;
{\bf(d)} as in (c) for sample SC.
}
\label{fig2}
\end{figure*}

Another important potential of LSPs lies in triggering coherent light emission amplified by plasmonic resonances. 
Coupling between a dye and the LSP can occur when the photoluminescence
spectrum of the semiconductor overlaps with the LSP resonance. Beyond a specific pump threshold, a resonant energy transfer between
the dye and the LSP can support lasing \cite{Gat12}. Examples of such systems are Spasers made of metal nano-particles 
embedded in a gain medium, \cite{Nog09,Sto08,Del11,Zhe08,Sto11,Liu11} plasmonic lasers and nano-cavities, \cite{Oul09,Rus12} random lasers with metallic scatterers. \cite{Pop06,Dice05}
Despite this cumulative effort in the field, the experimental demonstration of Spasers available in the literature are still very limited.
There is thus an opportunity to gain insight into the dynamics of surface plasmons supported lasing even in very simple systems. 
This in turn offers the possibility to investigate more complex lasing regimes, for example at pump energies that trigger higher order nonlinear effects.\\
Here we report on the lasing from systems based on polyhedral silver nanoparticles (nPs), with plasmonic response determined by
their shape and polydispersity. We show that a liquid dye solution in a parallelepiped cell acts as a laser once the metallic nPs with tailored plasmonic resonance are added: 
the system is hence a standard Fabry-Perot laser cavity, made by the facets of the cell and the gain is the dye solution, whose fluorescence 
is enhanced by the plasmonic resonances \cite{Nog06}. The lasing efficiency is strongly dependent on the resonance wavelength: 
the closer the resonance peak to the dye absorption maximum, the more efficient the process. Lasing is clearly demonstrated by the 
strong increase of the emitted radiation intensity and by the spectral narrowing through pump threshold. Moreover, we show that the 
lasing process evidenced by Fabry-Perot resonances of the cavity, gain saturation and non-resonant nonlinearities indicate that 
coherent emission can be observed in very dilute transparent samples, in contrast with random lasing, \cite{Wiersma95,Leonetti2011} relying on the
multiple scattering of light involving several particles in a volume. \\

\noindent
Of interest for both fundamental and applied sciences and applications, the synthesis of nPs has been associated 
with intense activities dedicated to further control the size and the shape of nanometer size colloids, soluble in a wide range of solvents. 
Metal nPs are subject to a wide range of investigations including optoelectronic devices, \cite{Ahn09,Rib11} sensing, 
\cite{Mah06,Tal10} catalysis, and plasmonics \cite{Gov07,Rig09,Tao07} to name but a few.  
In this work, the silver nP syntheses were carried out under ambient conditions according to modified protocols reported in the literature. 
Silver nitrate (AgNO$_3$, 99.0\% ACS grade), trisodium citrate (C$_6$H$_5$O$_7$Na$_3$.2H$_2$O, 99\% ACS grade), 
sodium borohydride (NaBH$_4$, 99\% ReagentPlus), polyvinylpyrrolidone (PVP, M$_{\textrm w} \sim$ 55 kD), 
L-Arginine (C$_6$N$_4$H$_{14}$O$_2$, 98\% reagent grade) were purchased from Sigma-Aldrich and used without further purification. 
For a typical preparation of spherical silver nPs, \cite{Yan05} silver nitrate solution (200~$\mu$L, 100~mM) was diluted in DI-water, 
the citrate solution (160~$\mu$L, 400~mM) was then added to act as a stabilizer for the nPs. Under vigorous stirring, 
a sodium borohydride solution (300~$\mu$L, 150~mM) was added drop-wise, initially leading to a black solution quickly 
turning into a yellow-brown solution. If stirring is stopped before the ageing is completed then part of the nPs falls out of solution. 
To address this effect and to obtain very stable colloidal solutions, the syntheses were left under stirring to age for at least 24 hrs. 
The total volume of the nPs solution was set to 20 mL but can easily be scaled up. Photoreduction has become a popular pathway 
to produce nPs of controlled shape including prism, disc and decahedral nPs. \cite{Jin01,Jin03,Xue07,Xue08,Zha10,Mai03,Wu08,Bas06,Cal03,Kim02,Nii03} Starting from small spherical nPs, varied 
light exposure conditions allow the preparation of nPs with tunable decahedron and prism populations. \cite{Pietrobon} 
In a typical synthesis, citrate solution (127~$\mu$L, 250~mM), PVP (19~$\mu$L, 50~mM), L-arginine (16~$\mu$L, 20~mM) 
and silver nitrate solution (25~$\mu$L, 50~mM) are successively mixed into DI-water. Sodium borohydride solution (102~$\mu$L, 100~mM) 
is swiftly injected leading a yellow pale solution which was exposed to light sources. The exposure conditions varied from 24 h to 2 weeks, and from monochromatic to white lights. 
Extraction by ultra-centrifugation was used to purify the as-prepared nPs solutions. 
Transmission electron microscopy (TEM) images were recorded using a Gatan CCD camera on a JEOL JEM-2011 electron microscope operating at 200 kV. 
Rhodamine B and the nPs aqueous solutions were mixed with the final dye concentration set to 1~mM for all the samples. 
A fraction of the resulting solutions were transferred in parallelepiped quartz cuvettes cells with a 1~mm optical path, 
with the water solution acting as gain medium once the system is pumped. A Coherent Surelite Q-switched frequency doubled 
Nd:YAG laser at wavelength $\lambda$=532~nm (repetition rate 10~Hz, pulse duration 6 ns and with 8 mm beam waist) 
was used for optical pumping. For each sample, the laser was focused down to a spot of 1 mm lateral dimension. 
The emitted radiation was collected from the input face, after focalization into an optical fiber connected to a 
Horiba MicroHR spectrograph equipped with Symphony electrically cooled CCD array detector, with gratings density of 600~mm$^{-1}$ and 1800~mm$^{-1}$, for low and high resolution spectra respectively.
The exposure time was set to $1$~s and each spectrum results from the average of 10 pump shots.

\begin{figure}
\includegraphics[width=0.45\textwidth]{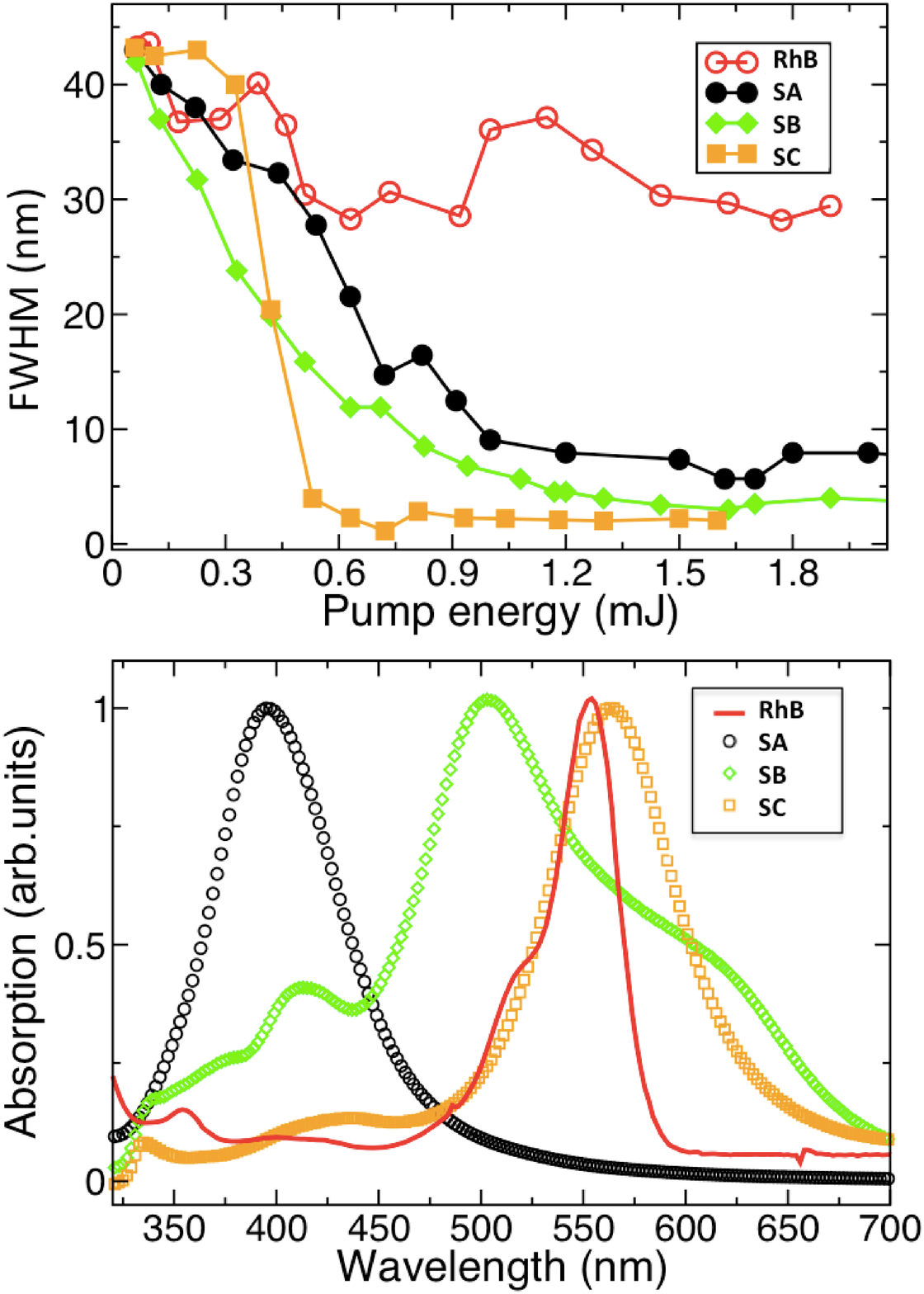}
\caption{ {\bf (a)} 
spectral width, calculated as FWHM of the emission for samples SA, SB, SC and pure Rhodamine solution  (open circles) when increasing input energy;
{\bf(b)} Absorption spectra of samples SA, SB, SC and pure RhodamineB solution  (thick red line). 
}
\label{fig3}
\end{figure}

\noindent

Figure \ref{fig1}a presents a selection of the samples used in this work. 
The different colors correspond to nPs with varied polyhedral morphologies, obtained by tuning the synthetic parameters. 
The transmission electron microscopy (TEM) images of two examples are shown in Figure \ref{fig1}b-c, for decahedral and prismatic shapes, respectively. 
The detailed population diagrams are presented in the 
Supporting Information, for the lasing experiments we focused mainly on the most representative samples SA, SB and SC.

\begin{figure}
\includegraphics[width=0.45\textwidth]{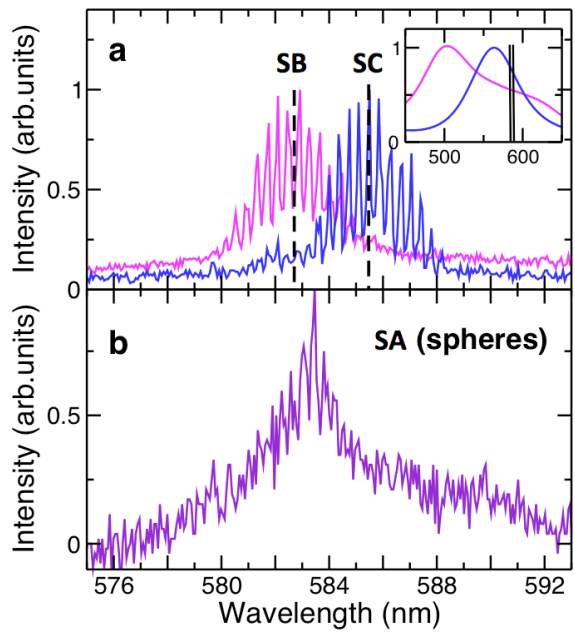}
\caption{{\bf(a)} High resolution emission spectra of samples SB and SC; the inset shows the corresponding absorption 
spectra of the nPs, with the vertical lines corresponding to the position of laser emission. {\bf(b)} Spectra of nanospheres sample (SA). }
\label{fig4}
\end{figure}

Figure \ref{fig2} shows the measured laser emission from these samples: by increasing the input pump energy the fluorescence 
spectrum of Rhodamine-B displays a pronounced peak that progressively becomes sharper and more intense, 
as reported in the panel a and b for samples SA with ellipsoidal nPs and SC with anisotropic and polyhedral nPs, respectively. 
In the same graphs the pure dye emission at much higher energy (2.5 mJ) is depicted showing that without nPs no lasing occurred (until being exposed to at least 4.5 mJ, not shown). 
This evidences clearly that lasing is provoked by the enhancement of dye fluorescence by the LSP resonances of the metallic colloids. 
Figure \ref{fig2}c-d presents the spectral width and the relative intensity of the emission as a function of the pump energy, 
unveiling the transition to a coherent emission. 
These results show that different samples have different lasing efficiency as evidenced by the spectral narrowing and by the saturation 
of the emission in SC, corresponding to a trend of the peak intensity deviating from a straight line as shown in Figure \ref{fig2}d.\\
To understand the reason of such dependency we report in Figure \ref{fig3}a the linewidth vs. pump energy for 
samples SA (spheres), SB, SC and pure Rhodamine solution, with a line-width of $\sim$3 nm for the polyhedrals and $\sim$10 nm for the spheres. 
In contrast, the control pure Rhodamine solution has a line width around $\sim$30 nm, in the same range of pump energy.
 Figure \ref{fig3}b shows the normalized absorption spectra of the plasmonic samples together and the pure dye. 
The resonance located around 400 nm  is characteristic of spherical silver nPs and presents little variation with the size of the nPs. \cite{Hao04} 
However, lower energy resonances are associated with silver nPs of a varied aspect ratios and dimensions. \cite{Moc02} 
The data demonstrate that samples with plasmonic resonance wavelength close to the dye absorption peak are more efficient lasers as the emission line is narrower. 
The different observed efficiency cannot be ascribed to the nPs concentration, being the concentration of the samples: [SA] = 17 nM, [SB] = 250 nM and [SC] = 14 nM. 
The low concentrations also evidence that the reported lasers are not random lasers, because the systems are totally transparent and there is no multiple light scattering.

\begin{figure}
\includegraphics[width=0.45\textwidth]{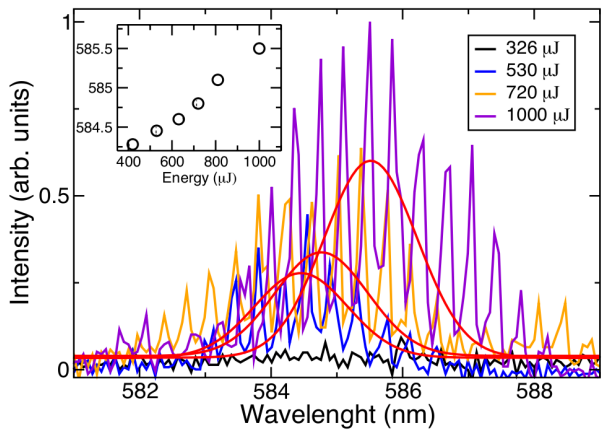}
\caption{Experimental emission spectra of SC - Rhodamine B solution collected over a range of input energy; solid lines are guides for the eye. Inset: envelop peak position vs. input energy}
\label{fig5}
\end{figure}

To gain further insight into the spectral emission we show in Figure \ref{fig4}a and Figure \ref{fig4}b the high resolution emission spectra of 
the samples respectively with polyhedral nPs (SB and SC) and with spherical nPs (SA), for pump energy of 1~mJ. 
The results require further analysis. We associate the emission spectra with the lasing modes of the Fabry-Perot cavity (the glass cuvette), 
hosting an active medium whose effective gain is enhanced by the metallic nPs. This claim is supported by the following evidences:\\
{\bf i)} The first observation is that, independently on the sample, the position of the maximum of the emission spectra approximately coincides with the emission peak of the Rhodamine ($\sim$ 580~nm). 
The absolute position shifts with the sample under consideration, likely due to a slightly different effective overlap between the absorption spectrum of the nPs and that of the dye. 
The inset in Figure \ref{fig4}a further highlights both these aspects.\\
{\bf ii)} The spacing of the spectral features is $\Delta\lambda\sim$0.5 nm for all the samples with polyhedral particles considered. 
This value has the same order of magnitude of the expected spacing $\Delta\lambda\approx\lambda^2 / $(2nL) = 0.13~nm of a 
Fabry-Perot cavity for normal incidence, with n~=~1.33 (Water), L~=~1mm and $\lambda$~=~532~nm. The difference between the experimental 
value and the model are most likely due to the resolution of our experimental setup (0.3~nm) and to a different effective index of the active medium.\\
{\bf iii)} The fringes are only visible for the samples with polyhedral nPs at this pump energy level, as evidenced in Figure \ref{fig4}b and from the
high resolution  emission spectra of pure RhodamineB-water solution reported in the Supplementary Information. 
This result reinforces the experimental findings of Figure \ref{fig3}, 
where nPs with absorption peak close to that of the dye yield to more efficient lasing. Control experiments with water alone confirm the lack of lasing. \\
A calibration experiment with Rhodamine only aqueous solutions (no nP present, see Supplementary Information) shows that lasing occurs at a threshold of $\sim$4.5~mJ. 
Assuming as lasing threshold the point of abrupt change of the slope of the emission peak intensity vs. pump energy in
Figure \ref{fig2}c-d, we can extrapolate the effective gain of the nP-dye solution. 
For example, for SC we experimentally obtain approximately a ten times increase in the effective gain, with respect to the dye alone, in the range of $\sim10^5~$cm$^{-1}$ 
(assuming that a 0.01~mM Rhodamine B solution in diethylene-glycol at 0.5~nJ/$\mu$m$^2$ has a gain of ~100 cm$^{-1}$).~ \cite{Leo2012}
This estimate assumes that the emission is the super-positioned output of many nanoparticles excited simultaneously. 
Thus variations in particle size and shape are expected to blur out any regular multi-mode emission spectrum.\\
The last aspect to analyze is the emission dynamics in the high pump energy range. From Figure \ref{fig2} it emerges that exceptionally, the lasing intensity only saturates for SC, 
which is the sample with the highest effective gain. This is a strong indication that the concentration of electromagnetic
energy associated with the plasmonic resonances could result in optical Kerr effects characterized by a spectral shift towards longer wavelength, with increasing energy.  
To verify this hypothesis we recorded several high-resolution spectra of the emission for sample SC, for increasing pump energy, as shown in Figure \ref{fig5}. 
The experimental results illustrate how increasing the input energy induces a red-shift of the emission shown in the inset of Figure \ref{fig5}.
This is consistent with our hypothesis and demonstrates the general relevance of non-resonant nonlinearities in plasmonic enhanced laser emissions.
It should be noted that the pulsed regime excludes that the shift could be due to thermal effects.
In the present experiments, as the resonances shift with increasing energy, their position with respect to the maximum (absorption) gain changes and this negative feedback results into a limited amplification. 
Remarkably, Figure \ref{fig5} also shows that the fine spectral resonances emerge nonlinearly from a spontaneous emission background, 
as they get more pronounced and visible by increasing pump. 
This is another evidence of the connection between the fine Fabry-Perot lasing modes and the laser efficiency, both boosted by the presence of resonant nPs. 

In conclusion we have unveiled novel aspects of the rich dynamics of plasmonic enhanced dye lasers, 
and specifically the pronounced non-linear regimes that can be achieved. Highlighting extremely strong light-matter interactions, 
these novel features include gain saturation due to Kerr effect. These results pave the way to the exploration of novel fundamental 
physical phenomena relative to nonlinear regimes at a spatial scale much smaller than the wavelength, 
as well as to innovative applications such as novel kind of laser sources providing ultra-short pulsed operation and 
structural tunability, realization of ultrafast microprocessors working at THz clock speed, supercontinuum generation, ultrasensing, ultradense and ultrafast information storage.

\section{Associated Content}
\subsection{Supporting Information}

Figure Supplementary1  reports population distribution of samples SA, SB and SC.
Figure Supplementary2  shows high resolution 
emission spectra of the aqueous Rhodamine-B solution used as gain material and the recorded peak intensity vs. pump energy.

\section{AUTHOR INFORMATION}
\subsection{Corresponding Authors}

\noindent
neda.ghofraniha[AT]roma1.infn.it\\
\noindent
pascal.andre[AT]st-andrews.ac.uk

\section{Acknowledgements}
The research leading to these results has received funding from the European Research Council under the European Community's Seventh Framework Program (FP7/2007-2013)/ERC grant agreement n.201766, project Light and Complexity and from the Italian Ministry of Education, University and Research under the Basic Research Investigation Fund (FIRB/2008) program/ CINECA grant code RBFR08M3P4. We thank MD Deen Islam for the technical support in the laboratory. P.A. would like acknowledge support of the Royal Society (RG080158) for the nP growth reactor design and fabrication, and the Scottish University Physics Alliance (SUPA) as well as the Canon Foundation in Europe to support his Advanced Research Fellowship at the University of St Andrews (UK) and his visiting scientist position at the RIKEN (Japan), respectively. A.D.F. is supported by an EPSRC Career Acceleration Fellowship (EP/I004602/1).



\end{document}